\documentclass[a4paper]{article}  
\usepackage{graphicx}

\begin{document}
	
\begin{center}
    {\Large Simulation of charmed particle production in Geant4}
\end{center}

\begin{center}
A. Galoyan$^{1}$,
V. Grichine$^{2}$,
A. Ribon$^{3}$,
V. Uzhinsky$^{4}$\\
\vspace{0.5cm}
on behalf of the Geant4 hadronic physics working group
\end{center}

\begin{center}
    \begin{minipage}{10cm}
$^{1}$VBLHEP, JINR, Dubna, Moscow region, 141980 Russia\\
$^{2}$P.N. Lebedev Physical Institute of RAS, 119991 Moscow, Russia\\
$^{3}$CERN, CH-1211 Genève 23, Geneva, Switzerland\\
$^{4}$LIT, JINR, Dubna, Moscow region, 141980 Russia
    \end{minipage}
\end{center}
\begin{center}
	\begin{minipage}{12cm}
	It is expected that charmed particles will be copiously produced at Future Circular Collider (FCC). 
	Due to relatively large life-time of the particles, it will be needed to account their interactions 
	with surrounded materials and detector’s materials. In order to satisfy the requirements, charmed particle 
	production in soft interactions has been implemented in QGS and FTF models of the Geant4 package. Some 
	details of the implementations are considered in the paper. 
    \end{minipage}
\end{center}

Charmed particles have been discovered and observed in fixed target experiments at accelerators of CERN, FERMILAB and DESY with hadronic beams at energies from 200 up to 900 GeV. In particular, they have been studied at SVD experiment in Protvino (Russia) with proton beam 70 GeV. Studies of the particles are included in work programs of large collaborations of RHIC and the LHC. Investigation of the charmed particles is one of the main tasks of future experiments at the accelerator complex NICA (Dubna, Russia) ‒ MPD and SPD. Copious production of the charmed particles is expected at the Future Circular Collider (FCC). Thus, it was decided to implement in the Geant4 package [1-3] production and transportation of the charmed particles which are available starting from Geant4 10.7 version.

Geant4 contains two main components ‒ simulation of the electro-magnetic interactions and simulation of strong hadronic collisions. The electro-magnetic part is well described in the papers [1-3].  Two models ‒ the Quark-Gluon String model (QGS) [4,5] and the Fritiof model (FTF) [6,7], are used for simulations of the hadronic interactions at high energies (E $>$ 3 GeV). For simulations of interactions with various materials, it is needed, first of all, to set up inelastic cross sections of hadrons with different nuclei. Here we use the so-called  Barashenkov-Glauber-Gribov cross sections (BGG) [3]. For calculations of interaction cross sections of the charmed particles, the approach described in Ref. [8] is used. Each of the models (QGS and FTF) use different methods for determining the multiplicity of intra-nuclear collisions. The FTF model applies the Glauber approximation and parameterizations of interaction cross sections of various hadrons with nucleons (see Ref. [3], p. 22, 23).  The procedure of the cross section calculations in the QGS model is well known [4,5,9]. At the ends of the stages, defined sets of quark-gluon strings are formed.

It is assumed in the FTF model that all inelastic hadron-hadron collisions are binary reactions: a) $h_1\ +\ h_2 \rightarrow h^*_1\ +\ h_2$; b) $h_1\ +\ h_2 \rightarrow h_1\ +\ h^*_2$; c) $h_1\ +\ h_2 \rightarrow h^*_1\ +\ h^*_2$. The process “a” is diffraction dissociation of a projectile hadron. The process “b” is diffraction dissociation of a target hadron. The process “c” represents non-diffractive processes. Experimental data are used for determination of cross sections or probabilities of the processes. $h^*_1$ and $h^*_2$ are excited states of the primary hadrons $h_1$ and $h_2$. The excited states are characterized by a mass $M_h$.  $M_h^2$ distribution in the processes “a” and “b” has the form $1/M_h^2$. The distribution in the process “c” is $D/M_h^2/\ln(M_{h,\ max}^2/M_{h,\ min}^2)\ +\ (1-D)/ (M_{h,\ max}^2- M_{h,\ min}^2)$ where $D=0.55$ is a parameter. The excited states are considered as quark-gluon strings, and they are subdivided into constituent anti-quark and quark (mesons) or quark and diquark (baryons). Observable hadrons are created by the string’s fragmentation. The fragmentation is carried out according to the LUND algorithm [10] in the case of the FTF model.

It is assumed in the QGS model that pomeron exchanges are dominating in the t-channel of elastic scatterings of hadron-hadron interactions at high energies. “Cutting” of the pomerons gives inelastic non-diffractive cross sections. Two quark-gluon strings are associated with each cut pomeron. Determination of the string masses and their kinematical properties are described in Ref. [11]. A special algorithm is used for the string fragmentation. Significant difficulties are the accounting of non-vacuum exchanges and the small mass string fragmentation in this approach.

Two processes are possible at a fragmentation of quarks or diquarks (anti-quarks or anti-diquarks): a creation of a meson ($M$) or a baryon ($B$) with corresponding probabilities -- $P_M$ or $P_B$ ($P_M=$ 93\% and $P_B=$ 7\% for quarks, and 
$P_M=$ 30\% and $P_B=$ 70\% for diquarks). It is assumed that these processes are happening at a creation of a quark-antiquark pair from vacuum (sea pair) in a field of the string with following pickup of quark or anti-quark by the fragmenting system. Production of pairs ‒ u–anti-u, d–anti-d, s–anti-s and c–anti-c, are possible with corresponding probabilities: 
$P_{u – anti-u}\ =\ P_{d – anti-d}\ =$\ 44 \%, $P_{s – anti-s} \simeq$ 12\% and $P_{c – anti-c} \simeq$ 0.02\%. The values of $P_{u – anti-u}$ and $P_{s – anti-s}$ were determined by comparing numerous calculations with various experimental data. The value of $P_{c – anti-c}$ was proposed in Refs. [12,13,14].

Having sampled a produced hadron, it is needed to determine its kinematical characteristics. First of all, the transverse momentum of the hadron, $P_T$, is determined. There are two methods for $P_T$ determination: usage of the gaussian distribution on $P_T$ -- $e^{- P_T^2/< P_T^2>}/\pi <P_T^2>$, or usage of distribution on transverse mass -- $m_T=\sqrt{m_h^2+P_T^2}$, in the form $B\cdot e^{-B\cdot (m_T-m_h)}$, $B\simeq$ 200 MeV$^{-1}$. Here $<P_T^2>$ is the average value of transverse momentum squared, and $m_h$ is the hadron mass. The first traditional method is originated from the papers [15,10,16]. The second method started to be used more recently [16,17].  

Having $P_T$, the longitudinal momentum of the hadron (along string axis) is determined as $P_L=(z\ P_0^+\ -\ m_T^2/z P_0^+)/2$, where $P_0^+$ is momentum of the string on light-cone, $P_0^+=E_0+P_{L0}$. A distribution on $z$ is called fragmentation function. 

As the fragmentation function, the symmetric LUND function is used in the FTF model -- $F(z) \propto z^\alpha\ (1-z)^\beta\ \exp(-b\ m_T^2/z)/z$, at $\alpha=0$, $\beta=1$, $b=0.7$  GeV$^{-2}$. An analogous function in QGS has the form: $F(z) \propto z^\alpha (1-z)\beta$. Tables of values of $\alpha$ and $\beta$ for various quarks and diquarks (anti-quarks and anti-diquarks) and various hadrons are given in [18 – 21].  An essential parameter of these fragmentation functions is the intercept of reggeon trajectory on which $c-anti-c$ mesons are located – $\alpha_\Psi(0)$. Following the paper [21], we choose $\alpha_\Psi(0) \simeq -2.2$.

The mass of a string is decreased after an emission of a hadron. Upon reaching a certain mass value, the string is considered as a hadron out of the mass-shell, and a procedure of putting it on the mass-shell is performed, that is, the string is ascribed by the real mass of the hadron, and momenta of all produced hadrons are re-determined to satisfy the energy-momentum conservation law, or 2-particle decay of the string is simulated. The second approach is used in Geant4, namely before each fragmentation step the possibility of 2-particle decay is checked, the probability distribution of which has the form: $\exp(-a\ (M_{st}^2 – M_{min}^2))$, $a=0.66$ GeV$^{-2}$. $M_{min}$ is the minimal mass of the string which, in the simplest approach, is a sum of masses of lightest hadrons what can be produced.

The most essential parameter of FTF and QGS models for charmed hadrons is $P_{c-anti-c}\simeq 0.02$\%, which determines overall yield of the charmed mesons. We have checked that the chosen value of the parameter allows to describe known experimental data (see Fig. 1).      
\begin{figure}[hbt]
	\centering 
	\resizebox{6in}{2.9in}{\includegraphics{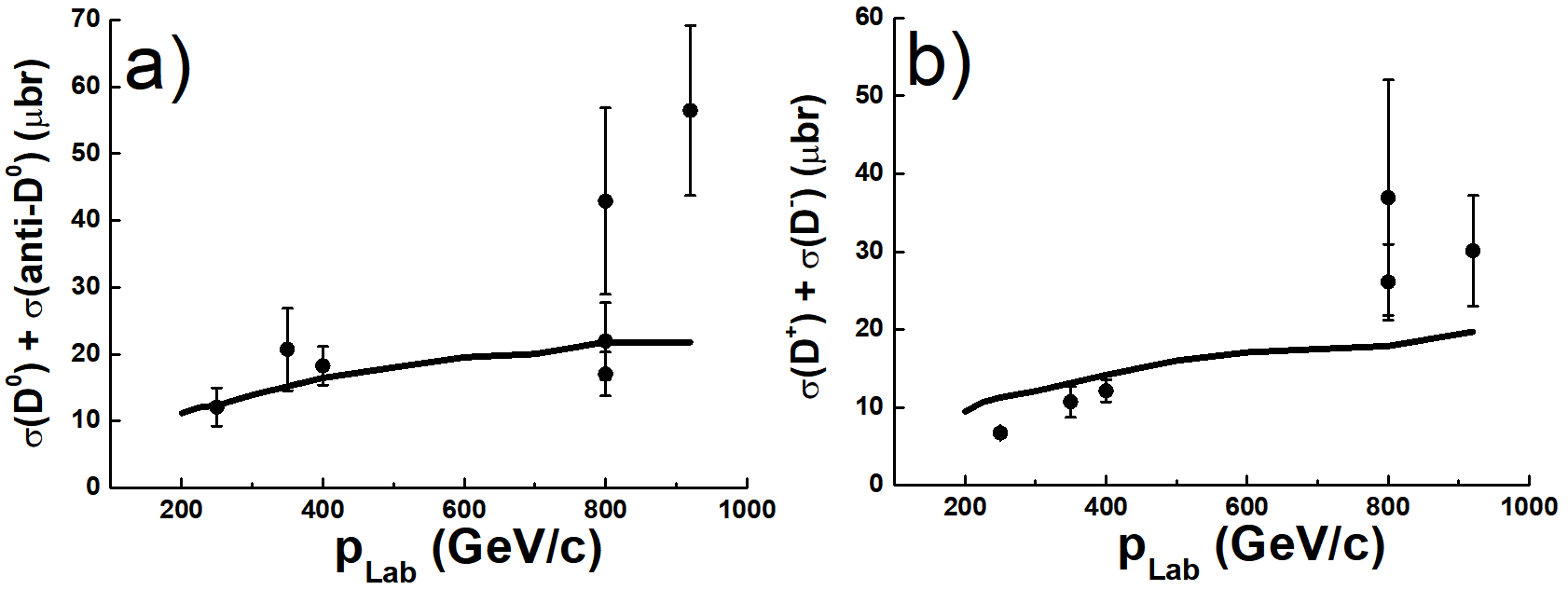}}
	\caption{Yields of $D^0$ and anti-$D^0$ mesons (fig. a), $D^+$ and $D^-$ mesons (fig. b) in $pp$-interactions as functions of momenta of projectile protons -- $P_{Lab}$. Solid lines are FTF model calculations. Solid points are experimental data presented in Ref. [22] (see Tabl. 4, p. 144).
	}
	\label{Fig1}
\end{figure}

The next important parameter in the QGS model is the average square of transverse momenta of hadrons ($<P_T^2>$) at the usage the gaussian distribution, or the parameter $B\simeq$ 200 MeV$^{-1}$ used in the $m_T$ distribution in the FTF model. By changing the $B$ parameter in FTF model, it is possible to describe $P_T^2$ distributions of the charmed particles up to $P_T^2\simeq$ 3 (GeV/c)$^2$ (see Fig. 2b). By using the gaussian distribution in the QGS model with $<P_T^2>\simeq$ 0.25 (GeV/c)$^2$, we fail to reproduce the $P_T^2$ distributions. Though, applying $m_T$-distribution in the QGS model one can obtain a satisfactory description (see Fig. 2d).
\begin{figure}[hbt]
	\centering 
	\resizebox{6in}{4in}{\includegraphics{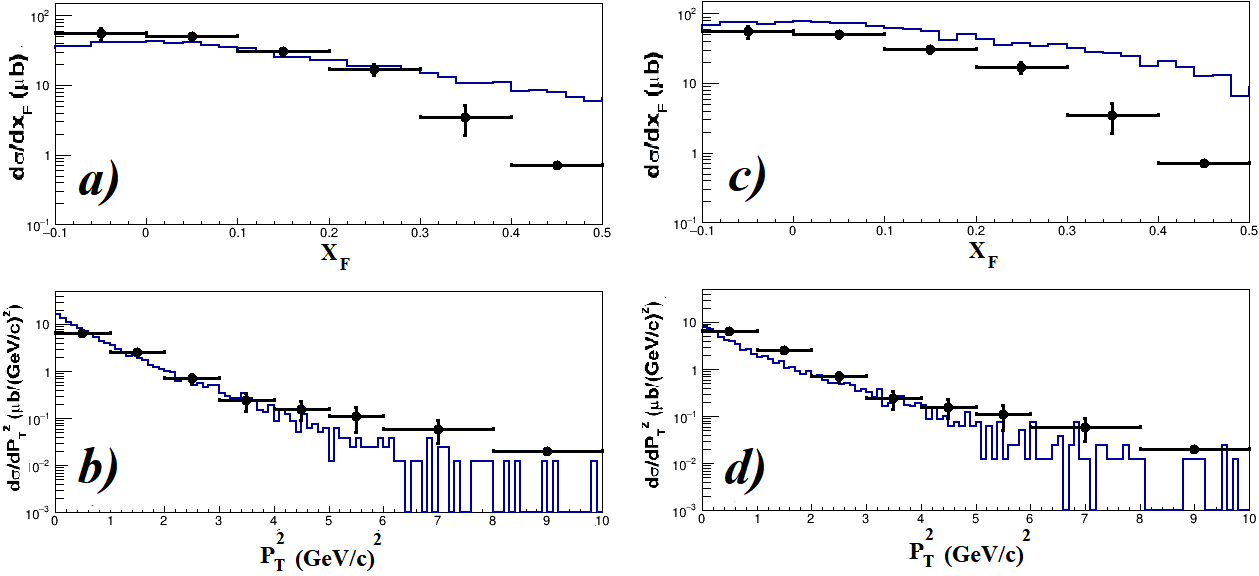}}
	\caption{$D^0$ and anti-$D^0$ meson distributions in $pp$-interactions at the momenta of projectile protons 250 GeV/c on $x_F$ variable and on squared transverse momenta in the FTF model (figs. a and b) and in the QGS model (figs. c and d). Points are experimental data of Ref. [23].  All calculations in FTF and QGS models are performed with the use of the $m_T$-distribution.
	}
	\label{Fig2}
\end{figure}

Hadron distributions on the Feynman $x_F$ variable are regulated by the fragmentation functions. As mentioned above, the LUND algorithm of the string fragmentation with symmetric fragmentation function is used in the FTF model. This allows to describe most of the known distributions of the charmed particles on $x_F$ (see Fig. 3). Though, at high energies, in particular at 920 GeV, we obtain a significantly reduced yield of $D^0$-mesons.  The most natural way out is to consider the “hard” hadron collisions described by the Quantum Chromo-Dynamics (QCD) which is not done up to now in the FTF Model.

\begin{figure}[hbt]
	\centering 
	\resizebox{6in}{4in}{\includegraphics{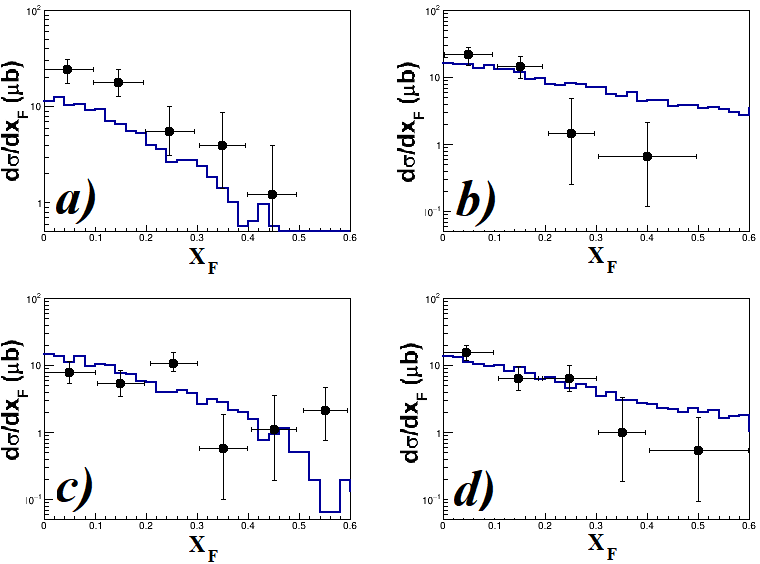}}
	\caption{$x_F$ distribution of $D^0$, anti-$D^0$, $D^+$ and $D^-$ mesons in $pp$-interactions at the momentum of the projectile protons 400 GeV/c (figs. a, b, c and d, respectively). Points are experimental data of Ref. [24]. The histograms are the FTF model calculation.
	}
	\label{Fig3}
\end{figure}

The ”hard” interactions are not also considered in the QGS model of the Geant4 package. 

\centerline{CONCLUSION}
Production of the charmed particles is implemented in the FTF and QGS models of the Geant4 package. Chosen values of the free parameters allow to reproduce existing experimental data except for the spectra of the “hard” particles. It is assumed that the charmed particles are mainly produced at quark-gluon string fragmentations.  With the calculation of cross sections of charmed particles with nuclei, it is possible to simulate the passage of these particles through matter. 

The authors are thankful to heterogeneous computer team of LIT JINR (HybriLIT) for support of calculations.

\end{document}